# Replacing CAPTCHA with XNO micropayments

Sujanavan Tiruvayipati

*Department of Computer Science and Engineering, Maturi Venkata Subba Rao Engineering College, Osmania University*
*Hyderabad, Telangana, India*
sujanavan_cse@mvsrec.edu.in

**Abstract:** As technology and gadgets continue to evolve, the need for bot-friendly and user-friendly internet becomes increasingly critical. This work discusses a methodology for implementation and feasibility of replacing traditional CAPTCHA mechanisms with Nano(XNO) cryptocurrency micropayments as a win-win solution and leverages the decentralized and secure nature of cryptocurrencies to introduce a micropayment-based authentication system. This approach not only enhances security by adding a financial barrier for automated bots but also provides a more seamless and efficient user experience. The benefits of this approach include reducing the burden on users while creating a socio-economic model that incentivizes internet service providers and content creators, even when accessed by bots. Furthermore, the integration of XNO micropayments could potentially contribute to the broader adoption and acceptance of digital currencies in everyday online transactions.

**Keywords**: Bot-friendly, Nano, XNO, Cryptocurrency, Blockchain, Block-lattice, CAPTCHA, Micropayments.

## I. INTRODUCTION

CAPTCHA, while effective in distinguishing between humans against bots, often leads to user frustration and prevents bots from accessing a server. Instead users or bots would make a small, virtually frictionless Nano(XNO) [1] [2] cryptocurrency payment to access the services provided by a server. XNO is designed to enhance scalability and does not involve any fees[3]. XNO utilizes a unique structure called the "Block Lattice", where each account has its own blockchain and allows for parallel processing of transactions, enhancing scalability and speed. XNO aims to provide near-instant transaction confirmation, ensuring that users do not have to wait for extended periods for their transactions to be processed. XNO is often praised [4] [5] for its energy-efficient design compared to traditional proof-of-work (PoW) cryptocurrencies like Bitcoin [6]. The absence of mining activities contributes to a negligible environmental footprint. XNO operates on a decentralized network, which means that no single entity has control over the entire network and reduces the risk of central points of failure. Anyone can get XNO from faucets [7] and all transactions on the XNO are transparent [8].

## II. PROPOSED METHODOLOGY

Following are the steps for implementing:-
1. Ask the user to connect a XNO wallet [9] address from which payment will be done.
2. Check if the wallet address is valid by querying the history of its transactions using a XNO public node [10] over RPC protocol [11].
3. To confirm the ownership of the wallet, ask the user to change the wallet representative.
4. Verify the representative change of the wallet by the user using a XNO public node.
5. Show the service provider's XNO payment information and ask the user for payment.
6. Validate Payment using a XNO public node along with recorded hash values of previous transactions and then provide the service to the user.

A demo was developed [12] to showcase the application of this proposed methodology.

## III. CONCLUSION

Introducing XNO micropayments for accessing a server would deter spammers and imbibe responsible usage among bot programmers, as they would have to incur a cost for each attempt. Websites, service providers and content creators could potentially earn revenue through micropayments for each completed task, providing an additional source of income. Users may be reluctant to adopt a system that requires payments, even if they are free and small. The idea of replacing CAPTCHA with XNO micropayments has potential benefits but, the success of such a system would depend on careful implementation and user education to ensure an inclusive online environment.

## REFERENCES

[1] C. LeMahieu, "Raiblocks," 2014.: https://github.com/clemahieu/raiblocks
[2] Nano(XNO) Living Whitepaper: https://docs.nano.org/living-whitepaper
[3] Nano | Eco-friendly & feeless digital currency: https://nano.org
[4] What is NANO coin: Facilitate Daily Payments At Zero Charge, 2021: https://phemex.com/academy/what-is-nano-coin
[5] Senatus, Why Nano may be the ultimate store of value and reserve currency, 2021: https://senatusspqr.medium.com/why-nano-is-the-ultimate-store-of-value-and-reserve-currency-3b0318844bc8
[6] Satoshi Nakamoto, Bitcoin: A Peer-to-Peer Electronic Cash System, 2008: https://bitcoin.org/bitcoin.pdf
[7] NanoLooker, Nano(XNO) Block Explorer: https://nanolooker.com
[8] Faucets for Nano(XNO): https://nanolooker.com/faucets
[9] Wallets for Nano(XNO): https://hub.nano.org/wallets
[10] Nano(XNO) Public Nodes: https://publicnodes.somenano.com
[11] RPC Protocol, Nano(XNO) Documentation: https://docs.nano.org/commands/rpc-protocol
[12] Pay Per Search: https://mvsrec.edu.in/pps, https://github.com/sujanavan/PPS